\documentstyle[emulateapj,epsfig]{article}
\input epsf           

\makeatletter

\newenvironment{inlinefigure}{%
\def\@captype{figure}%
\noindent\begin{minipage}{0.999\linewidth}\begin{center}}
{\end{center}\end{minipage}\smallskip}
\makeatother

\begin{document}
\title{The Redshift Evolution of the $2-8~$keV X-ray Luminosity Function}
\author{L.\,L.\ Cowie,$\!$\altaffilmark{1}
A.\,J.\ Barger,$\!$\altaffilmark{2,3,1}
M.\,W.\ Bautz,$\!$\altaffilmark{4}
W.\,N.\ Brandt,$\!$\altaffilmark{5}
G.\,P.\ Garmire$\!$\altaffilmark{5}
}

\altaffiltext{1}{Institute for Astronomy, University of Hawaii,
2680 Woodlawn Drive, Honolulu, Hawaii 96822}
\altaffiltext{2}{Department of Astronomy, University of Wisconsin-Madison,
475 North Charter Street, Madison, WI 53706}
\altaffiltext{3}{Department of Physics and Astronomy, University
of Hawaii, 2505 Correa Road, Honolulu, HI 96822}
\altaffiltext{4}{Center for Space Research,
Massachusetts Institute of Technology, Cambridge, MA 02139}
\altaffiltext{5}{Department of Astronomy \& Astrophysics,
525 Davey Laboratory, The Pennsylvania State University,
University Park, PA 16802}

\slugcomment{To appear in The Astrophysical Journal Letters}

\begin{abstract}
The high angular resolution and sensitivity of the 
{\it Chandra X-ray Observatory} 
has yielded large numbers of faint X-ray sources with measured
redshifts in the soft ($0.5-2$~keV) and hard
($2-8$~keV) energy bands. Many of these sources show few obvious 
optical signatures of active galactic nuclei (AGN). 
We use {\it Chandra} observations of the Hubble Deep Field North
region, A370, and the Hawaii Survey Fields SSA13 and SSA22, 
together with the {\it ROSAT} Ultra Deep Survey soft sample
and the {\it ASCA} Large Sky Survey hard sample, to construct
rest-frame $2-8$~keV luminosity functions versus redshift for all 
the X-ray sources, regardless of their optical AGN characteristics.
At $z=0.1-1$ most of the $2-8$~keV 
light density arises in sources with luminosities in the 
$10^{42}$~ergs~s$^{-1}$ to $10^{44}$~ergs~s$^{-1}$ range.
We show that the number density of sources in this luminosity 
range is rising, or is at least constant, with decreasing redshift.
Broad-line AGN are the dominant population at higher luminosities,
and these sources show the well-known rapid positive evolution
with increasing redshift to $z~\sim 3$.
We argue that the dominant supermassive black 
hole formation has occurred at recent times in objects with low 
accretion mass flow rates rather than at earlier times in more X-ray 
luminous objects with high accretion mass flow rates.
\end{abstract}


\section{Introduction}
\label{secintro}

Hard X-ray surveys provide the most direct probe of supermassive 
black hole (SMBH) accretion activity, since high energy X-rays can 
penetrate extremely large column densities of gas and dust.
Early analyses of deep {\it Chandra} X-ray surveys found
that optical selections of active galactic nuclei (AGN) had 
missed large numbers of accreting SMBHs.
\markcite{barger01b}Barger et al.\ (2001b) showed that the 
redshift history of SMBH growth is more strongly peaked 
to low redshifts than had been inferred from 
optically-selected samples.

With the high spatial resolution and energy sensitivity of deep 
{\it Chandra} observations, X-ray samples can be selected at 
both low and high redshifts in the same rest-frame {\it hard} 
energy band and their optical counterparts, and hence
spectroscopic redshifts ($\sim65$\% can be identified), unambiguously 
determined. The hardest band for which the hard X-ray luminosity 
function (HXLF) can presently be determined over a wide range of 
redshifts is rest-frame $2-8$~keV.
Although some Compton-thick sources may still be omitted,
the accounting will be far more complete than that of
any other available sample.

In this paper
we use the 1~Ms exposure of the {\it Chandra} Deep 
Field North (CDF-N) and three $\sim 100$~Ks
exposures of the A370, SSA13, and SSA22 fields, 
together with previous {\it ROSAT} and {\it ASCA} 
data, to construct the rest-frame $2-8$~keV luminosity function
from $z=0$ to 4 and model the SMBH accretion history.
We assume $\Omega_M=1/3$, $\Omega_\Lambda=2/3$, 
and $H_o=65$~km~s$^{-1}$~Mpc$^{-1}$. We use $L_x$ to denote
the rest-frame $2-8$~keV luminosity.

\section{X-ray Sample Selection}
\label{secsamp}

The faintest data are taken from the 1~Ms CDF-N catalogs of 
\markcite{brandt01}Brandt et al.\ (2001; hereafter B01)
and \markcite{barger02}Barger et al.\ (2002).
We exclude known stars and the small number of extended X-ray
sources. 
The flux limits are 
$\approx 3\times 10^{-17}$~ergs~cm$^{-2}$~s$^{-1}$
($0.5-2$~keV) and $\approx 2\times 10^{-16}$~ergs~cm$^{-2}$~s$^{-1}$
($2-8$~keV). 
Below a flux of $1.6\times 10^{-14}$~ergs~cm$^{-2}$~s$^{-1}$ ($2-8$~keV)
and $6\times 10^{-15}$~ergs~cm$^{-2}$~s$^{-1}$ ($0.5-2$~keV) we 
include only sources lying within $8'$ of the aimpoint where
the redshift identifications are most complete;
above these fluxes we include all the sources. 
This restricted sample contains 199 (52\%) sources in the $2-8$~keV band
and 219 (57\%) in the $0.5-2$~keV band, where the parentheses
are the percentages with redshifts.
We supplement the sample at fluxes above
$2\times 10^{-15}$~ergs~cm$^{-2}$~s$^{-1}$ ($2-8$~keV) and
$3\times 10^{-16}$~ergs~cm$^{-2}$~s$^{-1}$ ($0.5-2$~keV) with data 
from the SSA13 field (\markcite{barger01a}Barger et al.\ 2001a),
the A370 field (\markcite{barger01b}Barger et al.\ 2001b), and
the SSA22 field (A. J. Barger et al., in preparation).
This adds a further 67 (72\%) sources in the $2-8$~keV band
and 88 (60\%) in the $0.5-2$~keV band. We further supplement the 
sample with the larger area bright flux samples in the soft band
from {\it ROSAT} (\markcite{lehmann01}Lehmann et al.\ 2001)
and the hard band from {\it ASCA}
(\markcite{akiyama00}Akiyama et al.\ 2000). These have highly 
complete spectroscopic identifications. 

For the conversion of counts to flux in the CDF-N, we use 
the X-ray spectral photon indices $\Gamma$ determined in B01
from the hard-band to soft-band counts ratios. For
the conversions in the remaining {\it Chandra}
fields, we assume a fixed $\Gamma=1.2$, which is 
approximately the average $\Gamma$ in the B01 
sample. For the flux conversions in the {\it ROSAT} and {\it ASCA} 
samples, we use the softer $\Gamma$ values assumed by 
Lehmann et al.\ (2001) and Akiyama et al.\ (2000), which are 
appropriate for higher flux sources.

We determined
the solid angle covered by the sample at a given flux 
by comparing the observed numbers of objects as a
function of flux with the measured number counts in the 
appropriate energy band. For $2-8$~keV we use the 
averaged counts given in \markcite{cowie02}Cowie et al.\ (2002),
and for $0.5-2$~keV we use the power-law fits of 
\markcite{mushotzky00}Mushotzky et al.\ (2000), which agree 
with other determinations  
(e.g., B01; \markcite{rosati02}Rosati et al.\ 2002).
The solid angle covered by the combined
$2-8$~keV samples ranges from just under 0.01~deg$^2$ at
the faintest fluxes to 5.8~deg$^2$ at the highest fluxes.
At $10^{-14}$~ergs~cm$^{-2}$~s$^{-1}$ ($2-8$~keV) the solid
angle is 0.11~deg$^2$. The maximum solid angle covered by the 
combined $0.5-2$~keV samples is 0.42~deg$^2$. At 
$10^{-15}$~ergs~cm$^{-2}$~s$^{-1}$ ($0.5-2$~keV)
the solid angle is 0.12~deg$^2$.

%
%
\begin{inlinefigure}
\centerline{\epsfxsize=225pt\epsfbox{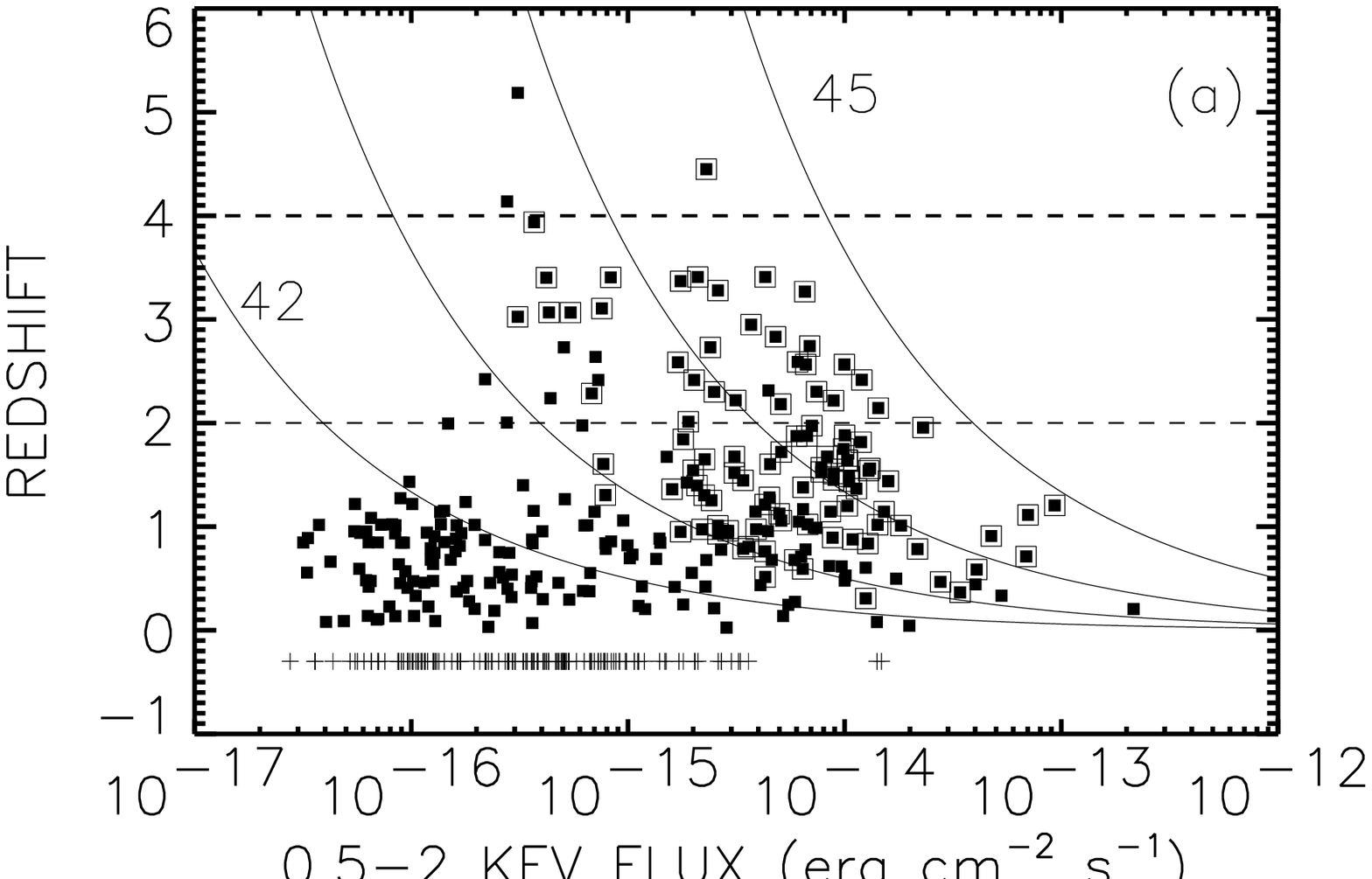}}
\centerline{\epsfxsize=225pt\epsfbox{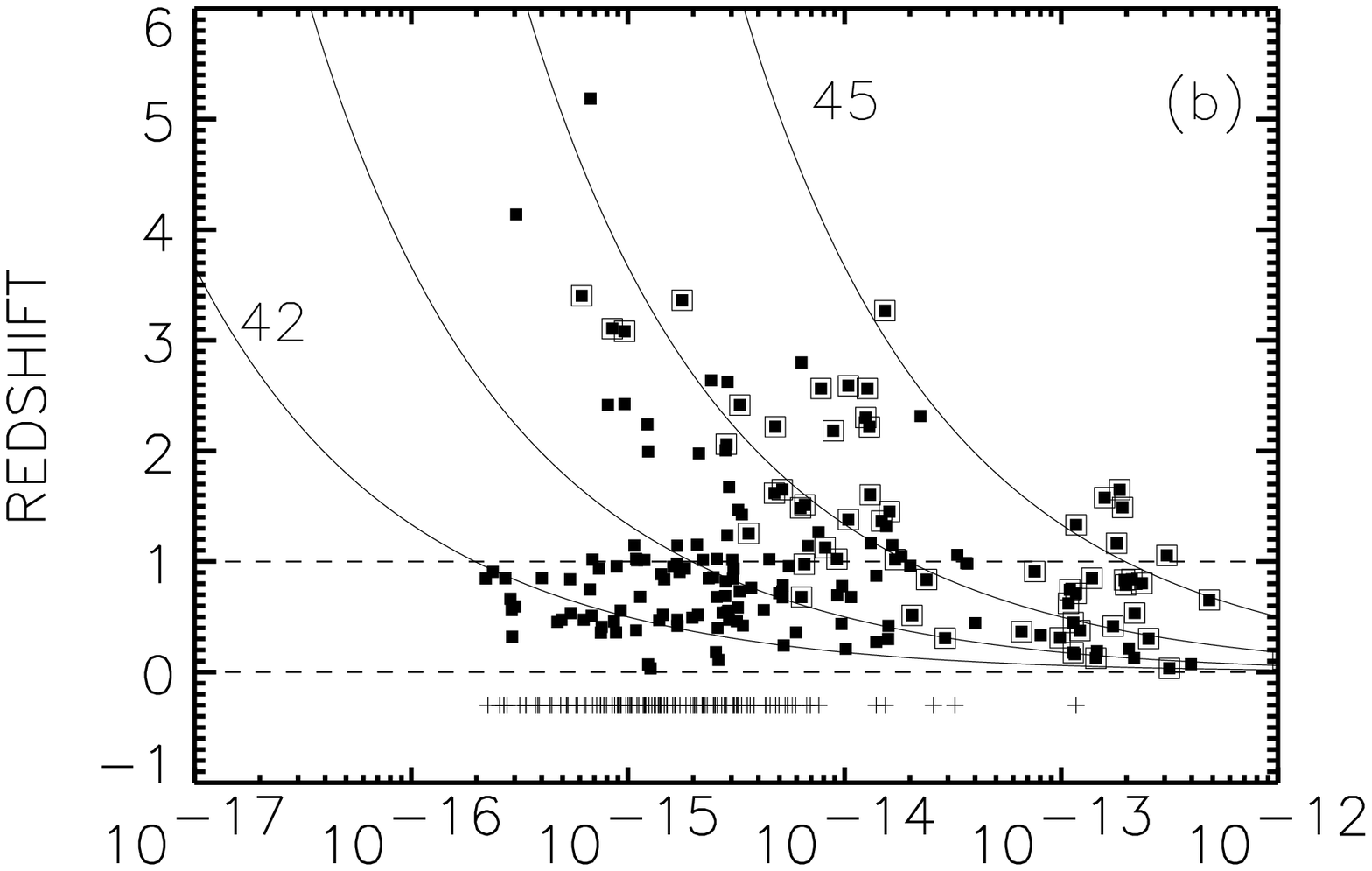}}
\vspace{12pt}
\figurenum{1}
\caption{
Redshift versus flux for sources selected in
(a) the $0.5-2$~keV band and (b) the $2-8$~keV band.
Solid curves show flux versus redshift for
sources with $L_x=10^{42}$~ergs~s$^{-1}$ (lowest curve),
$10^{43}$~ergs~s$^{-1}$, $10^{44}$~ergs~s$^{-1}$, and
$10^{45}$~ergs~s$^{-1}$ (highest curve),
computed with a $K$-correction for a $\Gamma=1.8$
power-law spectrum. Dashed lines show the redshift
intervals used in constructing the HXLFs.
Sources with broad-line optical spectra are
enclosed in a second, larger symbol. Spectroscopically
unidentified sources are denoted by plus signs at $z=-0.3$.
}
\label{fig1}
\addtolength{\baselineskip}{10pt}
\end{inlinefigure}

In Fig.~\ref{fig1}a and b we show redshift versus flux for the
soft and hard selected samples, respectively, including
sources that have not been spectroscopically identified.
The solid curves correspond to loci of constant $L_x$.
Sources with broad-line optical/UV spectra
are denoted by a second, larger symbol. Any source more luminous
than $10^{42}$~ergs~s$^{-1}$ is very likely an AGN on
energetic grounds
(\markcite{zezas98}Zezas, Georgantopoulos, \& Ward 1998;
\markcite{moran99}Moran et al.\ 1999),
though many of the intermediate luminosity sources
do not show obvious AGN signatures in the optical.
It is clear from Fig.~\ref{fig1} that the highest
X-ray luminosity sources are mostly broad-line AGN.

\section{Evolution of the AGN X-ray Luminosity Function}
\label{seclfs}

In the present section we concentrate on two redshift
intervals: $z=0.1-1$, as shown by
the dashed lines in Fig.~\ref{fig1}b, and
$z=2-4$, as shown
in Fig.~\ref{fig1}a. With these redshift intervals we can
compare samples chosen in approximately the same rest-frame hard
X-ray energy range of $3-12$~keV (the observed $2-8$~kev
band at the center of the low
redshift interval) and $2-8$~keV (the observed $0.5-2$~keV
band at the center of the high
redshift interval). The remaining small differential $K$-corrections
to rest-frame $2-8$~keV are made by assuming that
the spectra can be approximated by $\Gamma=1.8$ power-laws.
We have chosen
to use this spectral index since the differential K correction
is mostly significant in correcting from higher to lower
energies where the spectra may be better represented by
the unabsorbed intrinsic power law
(see \markcite{barger02}Barger et al.\ 2002).
However, the results are
only weakly sensitive to this assumption.
For a $\Gamma=1.2$ power-law K-correction the HXLF at $z=2-4$
is essentially unchanged while the HXLFs at $z=0.1-1$ shifts to
slightly lower luminosities (a factor of 1.27) while the shape
remains approximately the same.

 We have chosen not to correct
for the intrinsic absorption since this too requires
an assumption about the intrinsic spectrum and also about the
form of the absorption and any scattering. Since we are comparing
in the same rest frame at different redshifts this will not
affect the HXLF comparison  unless there is substantial evolution in the
intrinsic absorption. Moreover in the $2-8$~keV rest frame band
the average absorption corrections are not very large.
If we assume an intrinsic spectral index of $\Gamma=1.8$ and
a simple photoelectric absorption the average absorption correction
to $L_x$ is 1.14 in the $0.1-1$ redshift range and 1.53
in the $z=2-4$ range.

We probe to an $L_x$ below
$10^{43}$~ergs~s$^{-1}$ in the high redshift interval and 
below $10^{42}$~ergs~s$^{-1}$ in the low redshift interval.
Both samples are highly complete in redshift identifications
at high luminosities.
At $L_x>10^{44}$~ergs~s$^{-1}$ in the high redshift interval
there are 22 spectroscopically identified sources, 21 of which are 
broad-line AGN. At most there are a further 18 spectroscopically 
unidentified sources that could lie at these luminosities in this redshift
interval; however, a substantial fraction of these are likely at 
other redshifts. At $L_x>10^{44}$~ergs~s$^{-1}$ in the low redshift 
interval, 85\% of the identified sources are broad-line AGN, and 
there are only three unidentified sources that could lie at these
luminosities in this redshift interval. Thus, most of the high 
luminosity X-ray sources are broad-line AGN rather than obscured AGN. 

The differential HXLF as a function of X-ray luminosity and redshift,
$d\Phi(L_x, z)/{d\log L_x}$, is the number of X-ray sources per unit 
comoving volume per unit logarithmic luminosity. We adopt the $1/V_a$ 
method to estimate the binned HXLF in different redshift shells as
\begin{equation}
{{d\Phi}\over {d\log L_x}}=
{\sum_{i} {1/{V_{a}^i}}\over \Delta\log L_x} \,,
\end{equation}
where the $V_a^i$ is the total comoving volume in the redshift 
bin $\Delta z$ in which source $i$
with luminosity $L_x$ could lie and still be included in the 
sample and the summation is over all sources in the
given redshift interval and luminosity bin.
(Log denotes the base 10 logarithm.)

%
%
\begin{inlinefigure}
\centerline{\epsfxsize=225pt \epsfbox{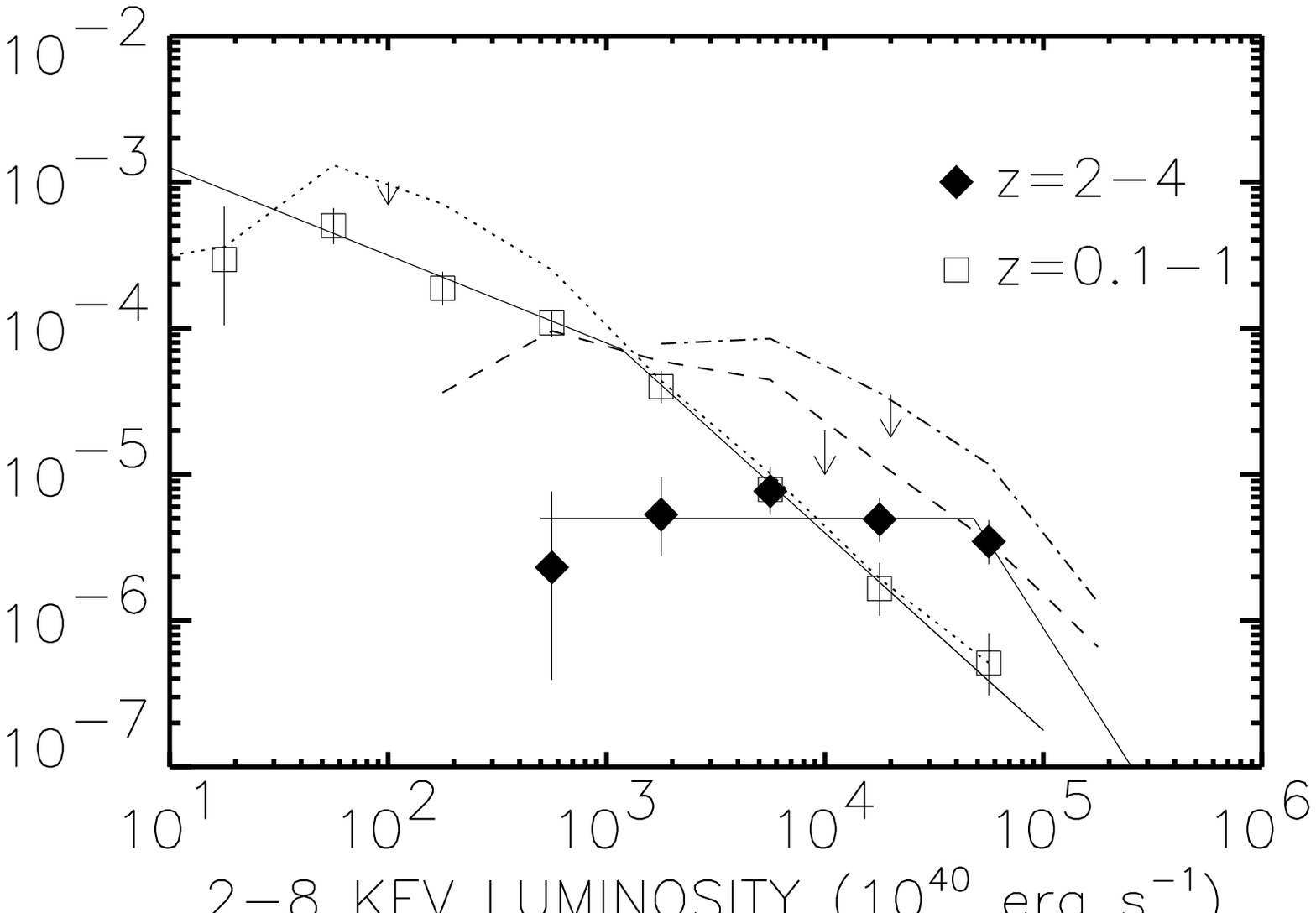}}
\vspace{6pt}
\figurenum{2}
\caption{
Rest-frame $2-8$~keV luminosity function
per unit logarithmic luminosity in the redshift
intervals $z=0.1-1$ (open squares) and $z=2-4$ (solid diamonds).
In the $z=0.1-1$ ($z=2-4$) interval the HXLF was
computed from observed-frame $2-8$~keV ($0.5-2$~keV).
An intrinsic $\Gamma=1.8$ was assumed, for which there is only a
small differential $K$-correction to rest-frame $2-8$~keV.
Poissonian $1\sigma$ uncertainties are based on the
number of sources in each luminosity bin.
Dotted ($z=0.1-1$) and dashed ($z=2-4$) lines are maximal
HXLFs found by assigning all the spectroscopically unidentified
sources to the center of each redshift interval.
Dot-dashed line is an upper bound on the $z=2-4$
HXLF obtained by selecting in observed-frame $2-8$~keV
and then assigning all the unidentified sources to this redshift
interval. Solid lines show a broken power-law fit to the
low redshift HXLF with a slope of $-0.6$ below
$1.2\times 10^{43}$~ergs~s$^{-1}$ and $-1.35$ above, and to
the high redshift HXLF with a slope of $0.0$ below
$4.8\times 10^{44}$~ergs~s$^{-1}$ and $-2.35$ above.
The upper $L_x$ fit in the high redshift interval
is based on the HXLF of Miyaji et al.\ (2001) shown in 
Fig.~\ref{fig3}.
}
\label{fig2}
\addtolength{\baselineskip}{10pt}
\end{inlinefigure}

Our measured HXLFs for the two redshift intervals are shown in
Fig.~\ref{fig2} with Poissonian $1\sigma$ uncertainties based
on the number of galaxies in each luminosity bin.
However, since incompleteness is a potentially larger source of
error, we recomputed the HXLFs by assigning all the
spectroscopically unidentified sources to the center of
each redshift interval (i.e., we included all the unidentified
sources in both redshift intervals).
Dotted ($z=0.1-1$) and dashed ($z=2-4$) lines in
Fig.~\ref{fig2} show these maximal limits.
Because the spectroscopic identifications are much more
complete at the high X-ray fluxes, the associated systematic
uncertainties are larger at lower $L_x$.
In principle photometric redshifts can be estimated for the
sources, and at higher redshifts this procedure is
effective (A. J. Barger et al., in preparation). However,
over the $z=0-4$ range there could be considerable
uncertainties in assigning photometric redshifts to sources with
substantial AGN contributions, so we prefer our
more conservative approach. Analyses of the HXLF using
alternate techniques suggest that the lowest luminosity  point
in the determinations
may be biased low (e.g. \markcite{page00}Page and Carrera 2000,
\markcite{miyaji01}Miyaji et al. 2001)
but the effect is small compared to the statistical and systematic
uncertainties.

At $L_x>10^{44}$~ergs~s$^{-1}$
Fig.~\ref{fig2} shows the familiar rise of the number counts 
with redshift. These counts are dominated by broad-line AGN. 
However, at lower $L_x$ the $z=0.1-1$ counts lie 
above the $z=2-4$ counts, or are comparable if we use the 
maximal correction for incompleteness. Since it is extremely 
unlikely that all the unidentified sources lie in the $z=2-4$ 
range (this would result in a large deficiency at $z=1-2$), 
the present-day number density of sources 
at $L_x\sim 10^{43}$~ergs~s$^{-1}$ is higher than it was
in the past. This result cannot be significantly changed by highly
obscured sources: the dot-dashed line shows the upper bound
on the $z=2-4$ HXLF if we selected 
in observed-frame $2-8$~keV, which corresponds
to rest-frame $8-32$~keV, and placed all the
unidentified sources in this redshift interval. The resulting 
upper limit is a factor of $\sim2$ higher than the upper 
limit obtained from the observed-frame $0.5-2$~keV selection.

All the X-ray sources 
are included in Fig.~\ref{fig2} without regard to optical 
spectroscopic classification.
Previous HXLF analyses have generally focused on 
broad-line AGN where the counterparts can be determined even in 
low resolution X-ray data (e.g., \markcite{boyle98}Boyle et al.\ 1998).
To make a direct comparison with these earlier studies, in Fig.~\ref{fig3} 
we plot our spectroscopically identified broad-line AGN HXLFs
in the intervals $z=0.1-1$ and $z=2-4$ with those derived
by \markcite{lafranca02}La Franca et al.\ (2002; $z=0.5$)
from {\it BeppoSAX} data and by
\markcite{miyaji01}Miyaji et al.\ (2001; $z=2-4$)
from {\it ROSAT} data.
Our HXLFs agree with these brighter HXLFs in the overlap 
region where there is also some overlap in the datasets used.
At $L_x\sim 10^{43}$~ergs~s$^{-1}$ there 
is little apparent evolution in the broad-line AGN HXLF, while at 
higher luminosities the HXLF increases with redshift.

%
%
\begin{inlinefigure}
\centerline{\epsfxsize=225pt \epsfbox{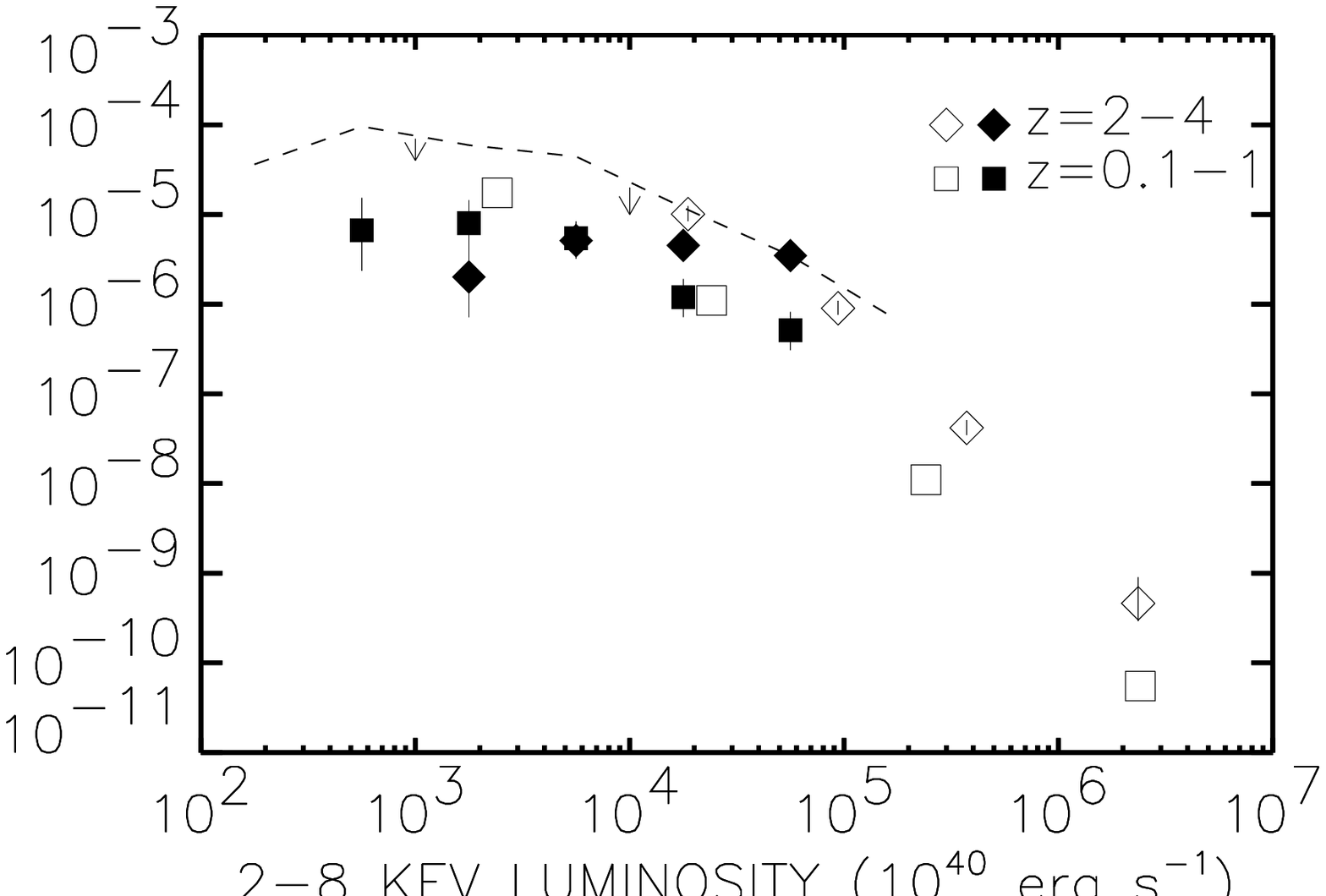}}
\vspace{6pt}
\figurenum{3}
\caption{
Rest-frame $2-8$~keV broad-line AGN luminosity function
per unit logarithmic luminosity
in the redshift intervals $z=0.1-1$ (solid squares)
and $z=2-4$ (solid diamonds). Dashed line is the maximal
HXLF in the $z=2-4$ interval found by assigning all the
spectroscopically unidentified sources to the center of the
interval. However, we expect that most of the broad-line AGN
have been identified since they are optically
bright and straightforward to identify spectroscopically.
Open boxes (open diamonds) show the broad-line AGN HXLF
at $z=0.5$ ($z=2-4$) derived by La Franca et al.\ (2002)
(Miyaji et al.\ 2001).
}
\label{fig3}
\addtolength{\baselineskip}{10pt}
\end{inlinefigure}

\section{Cosmic evolution}
\label{seccosmic}

In Fig.~\ref{fig4}a we show the comoving number density 
evolution with redshift of $L_x>10^{42}$~ergs~s$^{-1}$ sources.
At these luminosities the probability that the X-ray sources are
powered by anything other than AGN is extremely low, so we
are mapping the evolution of intermediate luminosity AGN.
In the $z=1-2$ redshift bin we show the number densities
from both observed-frame $2-8$~keV (open squares) and
observed-frame $0.5-2$~keV (solid diamonds); the results
are similar, although the soft band may be missing some obscured
sources. The solid ($2-8$~keV) and dotted ($0.5-2$~keV) horizontal 
bars show the number densities obtained if all 
the spectroscopically unidentified sources are assigned redshifts 
at the center of each redshift bin. The bars are not consistent with 
one another because all the unidentified sources are
included in all the bins.

%
%
\begin{inlinefigure}
\centerline{\epsfxsize=225pt \epsfbox{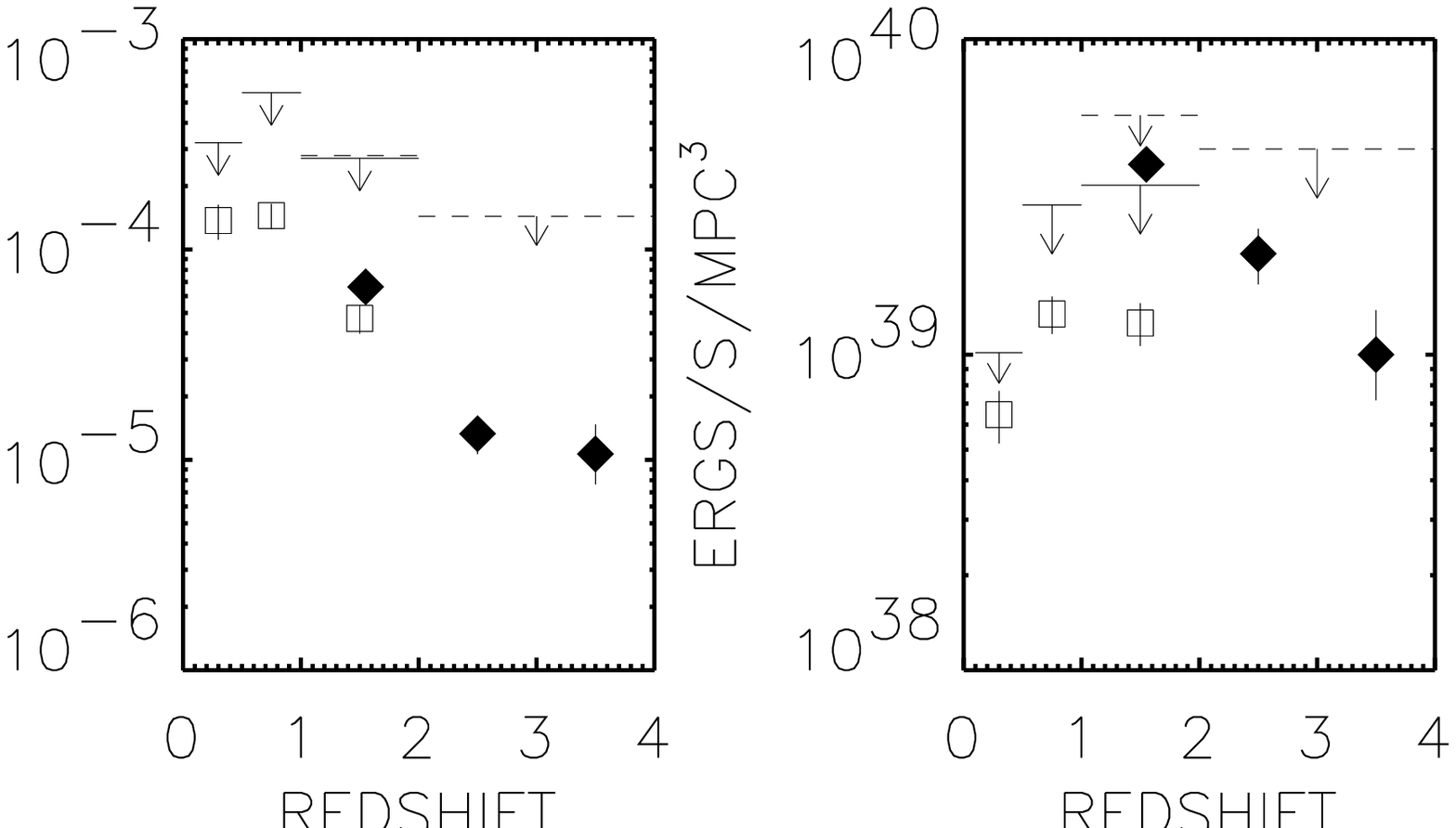}}
\vspace{4pt}
\figurenum{4}
\caption{
Evolution with redshift of the comoving number density
(a) and the $2-8$~keV comoving luminosity density
production rate
of $L_x>10^{42}$~ergs~s$^{-1}$ sources (b).
Open squares (solid diamonds) are measured from observed-frame
$2-8$~keV ($0.5-2$~keV). Poissonian $1\sigma$ uncertainties are
based on the number of sources in each bin. Dashed ($0.5-2$~keV)
and solid ($2-8$~keV) bars show upper limits found by assigning all
the unidentified sources to the center of each redshift bin,
so the bars are not consistent with one another.
To minimize this effect at high redshifts, we use
a single $z=2-4$ redshift bin. At the
highest redshifts a small correction (less than $20\%$) has
been applied to extrapolate from the minimum measured luminosity to
$L_x=10^{42}$~ergs~s$^{-1}$. Note that the uncertainty in the ${\dot\lambda_x}$,
which is weighted to higher luminosity sources,
is smaller than the uncertainty in the number density.
}
\label{fig4}
\addtolength{\baselineskip}{10pt}
\end{inlinefigure}

The number density of intermediate luminosity AGN 
does not rise rapidly from $z\sim 0$ to $z\sim 2.5-3$, 
in contrast to that of higher luminosity 
AGN (e.g., \markcite{miyaji00}Miyaji et al.\ 2000). 
In fact, the evolution with increasing
redshift is flat or declining. Thus,
AGN traced by X-ray luminosity show their own version of 
downsizing, just as star-forming galaxies do
(\markcite{cowie96}Cowie et al.\ 1996). 
High X-ray luminosity sources, which strongly overlap
with broad-line AGN (quasar) populations, peak at 
higher redshifts, while intermediate luminosity sources 
are still common now.

In Fig~\ref{fig4}b we show the evolution with
redshift of the $2-8$~keV comoving luminosity density production 
rate of $L_x>10^{42}$~ergs~s$^{-1}$ sources (${\dot\lambda_x}$). 
Integration of the power-law fits to the HXLFs shown
in Fig.~\ref{fig2} give a similar answer: 
$1.9\times 10^{39}$~ergs~s$^{-1}$~Mpc$^{-3}$ in the $z=0.1-1$
redshift interval and $2.2\times 10^{39}$~ergs~s$^{-1}$~Mpc$^{-3}$
in the $z=2-4$ interval. 
The largest uncertainty is the redshift distribution of the 
unidentified sources rather than the small effects of extrapolation outside
our observed luminosity range. 
This incompleteness uncertainty 
is less than a factor of 2.9 even in the high redshift 
interval (Fig.~\ref{fig4}b).

We parameterize the evolution of ${\dot\lambda_x}$
as

\begin{equation}
{{\dot\lambda_x}}= A~{\left(1+z\over 1.5\right)}^{\alpha} \,,
\end{equation}
where $\alpha$ ranges from
$-0.4$ (all unidentified sources in the low
redshift interval) to 1.1 (all unidentified sources in the
high redshift interval). If only identified sources
are included, ${\dot\lambda_x}$ is roughly flat until
the lowest $z=0.1-0.5$ interval in Fig.~\ref{fig4} where it
may drop slightly and $A\approx$~$2\times 10^{39}$~ergs~s$^{-1}$~Mpc$^{-3}$.
Thus, the $2-8$~keV luminosity density production 
rate evolves slowly with redshift, as was first noted in 
\markcite{barger01b}Barger et al.\ (2001b).

The SMBH accretion rate density is now given by
\begin{equation}
{{\dot\rho_{SMBH}}}= B~{{\dot\lambda_x}(1-\epsilon)\over \epsilon~c^{2}} \,,
\end{equation}
where $B$ is the correction from $L_x$ to the bolometric luminosity
of the AGN, and $\epsilon$ is the radiative efficiency
of the accretion flow. Setting $B~\sim~40$, which seems
to be roughly valid for both unobscured and obscured
AGN (\markcite{elvis94}Elvis et al.\ 1994; 
\markcite{barger01b}Barger et al.\ 2001b),
and $\epsilon~\sim~0.1$, which is close to the maximum possible 
radiative efficiency, 
\begin{equation}
{\dot\rho_{SMBH}}= C~{\left(1+z\over 1.5\right)}^{\alpha} \,,
\label{eqrhodot}
\end{equation}
where $C~\approx~1.3\times~10^{-5}$~${\rm M}_{\sun}$~yr$^{-1}$~Mpc$^{-3}$.
Even within the uncertainty in $\alpha$, SMBH growth is continuing
strongly to the present time. Integrating Eq.~\ref{eqrhodot} 
for $\alpha=0$ gives a present universal SMBH density of 
$\approx 2\times~10^{5}$~${\rm M}_{\sun}$~Mpc$^{-3}$,
comparable to the 
$\approx 2-4\times~10^{5}$~${\rm M}_{\sun}$~Mpc$^{-3}$
measured from local galaxies 
(\markcite{ferr00}Ferrarese \& Merritt 2000; 
\markcite{geb00}Gebhardt et al.\ 2000;
\markcite{yu02}Yu \& Tremaine 2002).
The late formation of the X-ray background reduces
the required local SBMH mass density of
previous estimates that assumed the bulk
of the production was earlier (\markcite{fab99}Fabian and Iwasawa 1999).

In summary, our results show that while the
higher accretion mass flow rates that power the most luminous
AGN peaked at higher redshift and are now much rarer,
lower accretion mass flow rates are still common at the present 
time. These less luminous events dominate SMBH formation.
The longer duration of the lower mass flow rate systems
provide constraints on theoretical modelling of the SMBH formation
(e.g., \markcite{haehnelt98}Haehnelt, Natarajan, \& Rees 1998;
\markcite{haiman00}Haiman \& Menou 2000;
\markcite{yu02}Yu \& Tremaine 2002) and on
the evolution of the accretion disks feeding the sources,
which may evolve more rapidly if they are initially of higher 
mass (Duschl 2002, private communication).

\acknowledgements
Support came from 
CXC grants DF1-2001X and GO2-3187B (LLC),
NSF grants AST-0084816 (LLC) and AST-0084847 (AJB),
the University of Wisconsin Research Committee with funds 
granted by the Wisconsin Alumni Research Foundation (AJB), 
the Alfred P. Sloan foundation (AJB),
NSF CAREER award AST-9983783 (WNB), and 
NASA grant NAS 8-01128 (GPG, PI).

\end{document}